\documentclass[a4paper, 12pt, openany, oneside]{article}
\usepackage[cp1251]{inputenc} %- для WiN
\usepackage[english,russian]{babel} %- для рус-англ. переноса
\usepackage{ graphics,amsfonts, amsmath,bm,amssymb, amsmath}
\usepackage[dvips]{graphicx}
\usepackage[flushleft,small]{caption2} % чтобы в подписях к рисункам
\renewcommand{\@biblabel}[1]{#1.\hfill}
\newcommand{\diag}{\rm \diag\, }

\renewcommand{\Re}{\mathop{\rm Re\,}}

\makeatother {

\begin{document}
 \thispagestyle{empty}
\large
\renewcommand{\refname}{\begin{center}\rm
REFERENCES%LITERATURE
\end{center}}

\begin{center}
SKIN EFFECT PROBLEM WITH DIFFUSION BOUNDARY CONDITIONS IN MAXWELL
PLASMA
\end{center}

\begin{center}
 {\bf Yu.F. Alabina, A.V. Latyshev, A.A. Yushkanov}
\end{center}

\begin{center}
 Moscow State Regional University\\
{\it 105005, Moscow, Radio st., 10 a\\
e-mail: yf.alabina@gmail.com, avlatyshev@mail.ru,
yushkanov@inbox.ru}
\end{center}

%\begin{abstract}
%We develope
The analytical method of solving the boundary problems for a system
of equations describing  the behaviour of electrons and an electric
field in the Maxwell plasma half-space is developed. Here the
diffusion reflection of electrons from the surface as a boundary
condition is supposed. The exact solution
 for boundary value problem  of  the Vlasov -- Maxwell
equations
 through the eigenfunctions of the discrete spectrum and the
 continuous
 spect\-rum is obtained.
The method of decomposition by eigenfunctions of discrete
 and continuous spectrum is used.
The exact
 expression for the surface impe\-dan\-ce also is obtained.\\
PACS numbers:52.35 - g; 52.20 - g; 52.25 - b\\

{\bf 1. Introduction. Statement of problem.}
 The skin effect is the
plasma's response to an external constant-amplitude variable
electromagnetic field that is tangential to the surface \cite{1}.
This problem represents essential interest \cite{1}--\cite{5},
with the focal point being on the computation of surface impedance.

In \cite{4} showed that electrons can transport the plasma
current away from the skin layer due their thermal motion
over distances of order $v_T/\omega$. In \cite{5} this
numerical study provides through specific numerical
comparisons a useful perspective on the limitations and
capabilities of the various Krook collisional models. This
information should be useful to future studies that try to
find a compromise in handling the yet unresolved problem of
hoe to unify the kinetic response and collisions in wave
phenomena.

    The paper further develops the analytical method of solving
boundary problem for a system of equations describing behaviour of
electrons and electric field in Maxwell plasma. The method is based
on the idea of the decomposition of the solution the problem
by singular
eigenfunctions of the corresponding characteristic equations.

We find the analytical solution to the boundary problem
of the skin
effect theory for electron plasma that fills the half-space. The
analytical solution has a form of the sum of an integral and
two (or
one) exponentially decreasing particular solutions of the initial
system. Depending on index the problem one of the particular
solutions disappears.

We consider a collision plasma with behaviour is described
by modelling kinetic equation with integral of collisions
in the form BGK--model
(BGK\-= \-Bhat\-na\-gar, Gross, Krook).

Let's Maxwell plasma fills the half-space $x>0$. Here $x$ is the
orthogonal coordinate  to the plasma boundary.
Let's the external
electric field has only  $y$ component. Then the self-consistent
electric field inside in plasma also has only $y$ component
$E_y(x,t)=E(x)e^{-i\omega t}$. We consider the kinetic equation
for the
electron distribution function:

$$
\dfrac{\partial f}{\partial t}+\text{v} _{x}\dfrac{\partial
f}{\partial x}+eE(x)e^{-i\omega t}\dfrac{\partial f}{\partial
p_y}=\nu(f_0-f(t,x,\mathbf{v})).
\eqno{(1.1)}
$$

In equation (1.1) $\nu$ is the frequency of electron
collisions with
ions, $e$ is the charge of electron, $f_0(\text{v})$ is the
equilibrium Maxwell distribution function, $\textbf{p}$ is the
momentum of electron,
$$
f_0(\text{v})=n\left(\dfrac{\beta}{\pi}\right)^{3/2}
\exp(-\beta^2\text{v}^2),\quad
\beta=\dfrac{m}{2k_BT}.
$$

Here  $m$ is the mass of electron, $k_B$ is the Boltzmann constant,
$T$ is the temperature of plasma, $\text{v}$ is the velocity of the
electron, $n$ is the concentration of electrons, $c$ is the
speed of
light.

The electric field $E(x)$  satisfies to the equation:
$$
E''(x)+\dfrac{\omega^2}{c^2}E(x)= -\dfrac{4\pi i  e^{i\omega
t}\omega e}{c^2} \int v_y f(t,x,\mathbf{v})\,d^3v. \eqno{(1.2)}
$$

We assume that intensity of an electric field is such that linear
approxima\-ti\-on is valid. Then distribution function can be
presented in the form:
$$
f=f_0\left(1+C_y\exp(-i\omega t)h(x,\mu)\right),
$$
where $\textbf{C}=\sqrt{\beta}\text{v}$ is the dimensionless
velocity of electron, $\mu=C_x$. Let $l=v_T\tau$ is the mean free
path of electrons, $v_T=1/\sqrt{\beta}$, $v_T$ is the thermal
electron velocity,\; $\tau=1/\nu$.
We introduce
the dimensionless parameters and the electric field:
$$
t_1=\nu t,\quad x_1=\dfrac{x}{l}, \quad
e(x_1)=\dfrac{\sqrt{2}e}{\nu\sqrt{mk_BT}}E(x_1).
$$

Later we substitute $x_1$ for $x$. The substitution produces the
following form of the kinetic equation (1.1) and the equation on a
field with the displacement current (1.2):
$$
\mu\dfrac{\partial h}{\partial x}+z_0\,h(x,\mu)=e(x),  \quad
z_0=1-i\omega\tau, \eqno{(1.3)}
$$
$$
%\dfrac{d^2e(x)}{d\,x^2}
e''(x)+Q^2e(x)=-i\dfrac{\alpha}{\sqrt{\pi}}
\int\limits_{-\infty}^{\infty}\exp(-{\mu'}^2)\,h(x,\mu')\,d\mu',
\quad
Q=\dfrac{\omega l}{c},
\eqno{(1.4)}
$$
where $\delta=\dfrac{c^2}{2\pi\omega\sigma_0}$,
$\delta$ is the classical
depth of the skin layer, $\sigma_0=\dfrac{e^2n}{m\nu}$,
$\sigma_0$ is the
electric conductance, $\alpha=\dfrac{2l^2}{\delta^2}$, $\alpha$ is
the anomaly parameter.

We formulate the boundary conditions for the distribution function
of the electron in case of the diffusion electron reflection
from the
surface:
$$
h(0,\mu)=0, \qquad 0<\mu<+\infty. \eqno{(1.5)}
$$
We use the condition that function $h(x,\mu)$ vanishes far from the
surface:
$$
h(+\infty,\mu)=0, \qquad -\infty<\mu<+\infty, \eqno{(1.6)}
$$
and conditions for electric field on the interface and far from it:
$$
{e}'(0)={e_s}', \qquad e(+\infty)=0, \eqno{(1.7)}
$$
where ${e_s}'$ is the given value of the gradient of the
electric field on the plasma interface.

So, the skin effect problem is formulated completely. We seek
solution of system of the equations (1.3) and (1.4)
in this problem,
that satisfy to the boundary conditions (1.5) -- (1.7).

 {\bf 2. Eigenfunctions and eigenvalues.} The separation of variables
\cite{2}--\cite{4}
$$
h_\eta(x,\mu)=\exp(-z_0\dfrac{x}{\eta})\Phi(\eta,\mu), \eqno{(2.1)}
$$
$$
e_\eta(x)=\exp(-z_0\dfrac{x}{\eta})E(\eta), \eqno{(2.2)}
$$
where $\eta$ is the spectral parameter (generally, complex-valued),
reduces  system of the equation (1.3) and (1.4) to the
characteristic system:
$$
(\eta-\mu)\Phi(\eta,\mu)=\dfrac{\eta}{z_0}E(\eta), \eqno{(2.3)}
$$
$$
\left[z_0^2+Q
 ^2\eta^2\right]E(\eta)=-\dfrac{i\alpha \eta^2}{\sqrt{\pi}}n(\eta),
\eqno{(2.4)}
$$
where
$$
n(\eta)=\int\limits_{-\infty}^{\infty}e^{-\mu^2} \Phi(\eta,\mu)d\mu.
$$

From equations (2.3) and (2.4) for $\eta\in(-\infty,\infty)$, we
find the eigenfunctions of the continuous spectrum \cite{2},
\cite{6}:
$$
\Phi(\eta,\mu)=\dfrac{a}{\sqrt{\pi}}\eta^3e^{-\eta^2}
P\dfrac{1}{\eta-\mu}+ \lambda(\mu)\delta(\eta-\mu), \eqno{(2.6)}
$$
$$
E(\eta)=\dfrac{az_0}{\sqrt{\pi}}\eta^2e^{-\eta^2}, \quad
a=-i\dfrac{\alpha}{z_0^3}. \eqno{(2.7)}
$$

In the equation (2.6) the symbol $Px^{-1}$ denotes the distribution,
i.e. the principal value of the integral of  $x^{-1}$, $\delta(x)$
is the Dirac delta function.

Since the distribution and electric field functions decline as we go
further from the boundary, we regard the continuous as positive real
half-axis:  $0<\eta<+\infty$. The eigen solution of the continuous
spectrum $h_\eta(x,\mu)$, $e_\eta(x)$ are decreasing from value $x$,
since $\Re z_0>0$.

In the equations (2.6) and (2.7) there is the normalization
condition:
$$
\int\limits_{-\infty}^{\infty}e^{-\mu^2}\Phi(\eta,\mu)d\mu=
\left[1+\left(\dfrac{\omega l}{c}\right)^2\eta^2\right]e^{-\eta^2},
$$
$\lambda(z)$ is the dispersion function,
$$
\lambda(z)=1+\left(\dfrac{Q}{z_0}\right)^2z^2+\dfrac{az^3}{\sqrt{\pi}}
\int\limits_{-\infty}^{\infty} \dfrac{e^{-\mu^2}d\mu}{\mu-z}.
\eqno{(2.8)}
$$

Consider  the dispersion function in the following form:
%$$
%\lambda(z)=1+bz^2+\dfrac{az^3}{\sqrt{\pi}}
%\int\limits_{-\infty}^{\infty}\dfrac{\exp(-\mu^2)\,d\mu}{\mu-z},
%\quad b=\dfrac{Q^2}{z_0^2}=\dfrac{(\omega\, l)^2}{c^2\Big(1-i
%\dfrac{\omega}{\nu}\Big)^2}.
%$$

%$$
%\lambda^+(\mu)-\lambda^-(\mu)=2\sqrt{\pi}\,i\,a\,\mu^3\,\exp(-\mu^2).
%\eqno{(2.19)}
%$$

$$
\lambda (z)=1+b\,z^2-a\,p(z),
$$
here \quad $b=\dfrac{Q^2}{z_0^2}=\dfrac{(\omega\, l)^2}{c^2\Big(1-i
\dfrac{\omega}{\nu}\Big)^2},\quad
p(z)=-\dfrac{z^3}{\sqrt{\pi}}\int\limits_{-\infty}^{\infty}
\dfrac{\exp(-\mu^2)\,d\mu}{\mu-z}.$

 $$ p^{\pm}(\mu)=p(\mu)\pm i q(\mu),\qquad
q(\mu)=\sqrt{\pi}\mu^3\exp(-\mu^2), $$
$$
b=\dfrac{\omega_1^2}{\nu_1^2\Big(1-i\dfrac{\omega_1}{\nu_1}\Big)^2}
\Big(\dfrac{v_T}{c}\Big)^2=
\dfrac{\omega_1^2}{(\nu_1-i\omega_1)^2}\Big(\dfrac{v_T}{c}\Big)^2,
$$
$$
a=-i\dfrac{\omega_1}{\nu_1^3\Big(1-i\dfrac{\omega_1}{\nu_1}\Big)^3}=
\dfrac{-i\omega_1}{(\nu_1-i\omega_1)^3.}
$$

The fact that $\lambda(\mu)$  has a double pole at $z=\infty$
follows from the asymptotic series    in neighborhood of infinity:
$$
\lambda(z)=(b-a)z^2+\Big(1-\dfrac{a}{2}\Big)-\dfrac{3a}{4}\cdot
\dfrac{1}{z^2}-\dfrac{15a}{8z^4}\cdot\dfrac{1}{z^4}\cdots.
\eqno{(2.13)}
$$

{\bf 3. Mode of plasma in Maxwell plasma.}    Find out the structure
of discrete spectrum, this spectrum consists of the zeros of the
dispersion function. Each zero corresponds to the own decision of
the discrete spectrum, named also mode of plasma.

Take two line $\Gamma_\varepsilon^{\pm}$, parallel real axis and
defending from it on distance$\varepsilon,\; \varepsilon>0$. The
value $\varepsilon$ shall choose so small that all zeroes to
dispersion function lay outside of narrow band, concluded between
direct $\Gamma_\varepsilon^{+}$ and $\Gamma_\varepsilon^{-}$.

According to principle of the argument difference between number of
the zeroes and number pole to dispersion function is an
incrementation of its logarithm:
$$
N-P=\dfrac{1}{2\pi
i}\Bigg[\;\int\limits_{\Gamma_\varepsilon^+}+
\int\limits_{\Gamma_\varepsilon^-}\Bigg]\,d\,\ln \lambda(z).
\eqno{(3.1)}
$$

In (3.1) each zero and pole are considered so much once as their
multiplicity, the lines $\Gamma_\varepsilon^{+}$ and
$\Gamma_\varepsilon^{-}$ are passed accordingly in positive and
negative directions. Then the dispersion function in infinitely
removed point has a pole of the second order, i.e. $P=2$.

When $\varepsilon\to 0$ from (3.1) we get:
$$
N-2=\dfrac{1}{2\pi i}\int\limits_{-\infty}^{\infty}\, d\ln
\dfrac{\lambda^+(\mu)}{\lambda^-(\mu)}. \eqno{(3.2)}
$$

Integral from (3.2) represented as:
$$
\int\limits_{-\infty}^{\infty} d\ln
\dfrac{\lambda^+(\mu)}{\lambda^-(\mu)}=
\int\limits_{0}^{\infty} d\ln
\dfrac{\lambda^+(\mu)}{\lambda^-(\mu)}+
\int\limits_{-\infty}^{0} d\ln
\dfrac{\lambda^+(\mu)}{\lambda^-(\mu)}.
$$

In the second integral we shall do change the variable: $\tau\to
-\tau$, we have:
$$
\lambda^+(-\tau)=\lambda^-(\tau), \qquad
\lambda^-(-\tau)=\lambda^+(\tau).
$$
Consequently, second integral is first. Really,
$$
\int\limits_{-\infty}^{0} d\ln
\dfrac{\lambda^+(\mu)}{\lambda^-(\mu)}=-
\int\limits_{0}^{\infty} d\ln
\dfrac{\lambda^+(-\mu)}{\lambda^-(-\mu)}=
-\int\limits_{0}^{\infty} d\ln
\dfrac{\lambda^-(\mu)}{\lambda^+(\mu)}=
\int\limits_{0}^{\infty} d\ln
\dfrac{\lambda^+(\mu)}{\lambda^-(\mu)}.
$$
Thereby,
$$
N-2=\dfrac{1}{\pi i}\int\limits_{0}^{\infty}\,d\ln
\dfrac{\lambda^+(\mu)}{\lambda^-(\mu)}. \eqno{(3.3)}
$$

On complex plane we consider line $\Gamma: \; z=G(t), \;0\leqslant t
\leqslant +\infty$, where $ G(t)=\dfrac{\lambda^+(t)}{\lambda^-(t)}.
$ Not difficult check that
$$
G(0)=1,\qquad \lim\limits_{\tau\to +\infty} G(t)=1.
$$

These equality mean that curve $\Gamma$ is closed: it comes out of
points $z=1$ and ends in this point. According to (3.3):
$$
N-2=\dfrac{1}{\pi i}\Big[\ln |G(\tau)|+i\arg G(\tau)\Big]
_0^{+\infty}=\dfrac{1}{\pi}\Big[\arg G(\tau) \Big]_0^{+\infty}.
$$
Thence we get:
$$
N-2=\dfrac{1}{\pi}\Big[\arg
G(t)\Big]_0^{+\infty}=2\varkappa(G), \eqno{(3.4)}
$$
where $\varkappa=\varkappa(G)$ is the index of the problem, $G(t)$
is the number of revolution the curve $\Gamma$ comparatively begin
coordinates, made in positive direction.

From (3.4) we see
$$
N=2+\dfrac{1}{\pi}\Big[\arg G(+\infty)-\arg G(0)\Big]=2+
\dfrac{1}{\pi}\arg G(+\infty), \eqno{(3.5)}
$$
since $\arg G(0)=0$.

$\arg G(t)$ is the regular branch of the argument, fixed in zero
condition: $\arg G(0)=0$, and determined in cut complex plane with
cut along positive part of real axis. This branch complies with the
main by importance of the argument.

We present the dispersion function in form (the displacement current
is absent):
$$
\lambda(z)=-ic\Big[\dfrac{i}{c}+\dfrac{z^3}{\sqrt{\pi}}
\int\limits_{-\infty}^{\infty}\dfrac{\exp(-\mu^2)\,d\mu}
{\mu-z}\Big]=-ic\omega(z),
%\eqno{(3.6)}
$$
where
$$
\omega(z)=\delta+\dfrac{z^3}{\sqrt{\pi}}
\int\limits_{-\infty}^{\infty}\dfrac{\exp(-\mu^2)\,d\mu}
{\mu-z},\qquad \delta=\dfrac{i}{c}. \eqno{(3.6)}
$$
According to (3.6) border values of the function $\omega(z)$
overhand and from below on real axis are: $
\omega^{\pm}(\mu)=\omega(\mu)\pm i \sqrt{\pi}\mu^3\exp(-\mu^2).
%\eqno{(3.8)}
$

Then, the $G(\mu)$ is equal:
$$
G(\mu)=\dfrac{\delta_1-p(\mu)+i[\delta_2+q(\mu)]}
{\delta_1-p(\mu)+i[\delta_2-q(\mu)]}. \eqno{(3.7)}
$$
Where
$$
\delta=\delta_1+i\delta_2=-\dfrac{4c^2}{\pi v_T^2 \omega_p^2}
\cdot\dfrac{(\omega+i \nu)^3}{\omega},
$$
whence
$$
\delta_1=-\dfrac{\omega^3-3\omega
\nu^2}{v_c^2\omega\omega_p^2},\quad
\delta_2=-\dfrac{3\omega^2\nu-\nu^3}{v_c^2\omega\omega_p^2},\quad
v_c=\sqrt{\dfrac{\pi}{4}}\dfrac{v_T}{c}.
$$

Select in equality (3.7) for function $G(\mu)$ real and imaginary
part: $ G(\mu)=G_1(\mu)+iG_2(\mu). $ We have
$$
G_1(\mu)=\dfrac{[\delta_1-p(\mu)]^2+\delta_2^2-q^2(\mu)}
{[\delta_1-p(\mu)]^2+[\delta_2-q(\mu)]^2},
$$
$$
G_2(\mu)=\dfrac{2\,q(\mu)\,[\delta_1-p(\mu)]}
{[\delta_1-p(\mu)]^2+[\delta_2-q(\mu)]^2}.
$$

On complex $\delta$--planes shall enter curve $\Lambda$:
$$
\Lambda: \qquad\delta_1=p(\mu), \quad \delta_2=\pm q(\mu),
\quad 0\leqslant \mu \leqslant +\infty.
$$

Entrails the curve $\Lambda$ shall mark through $\Delta^+$, and
$\Delta^-$ is the external of curve $\Lambda$. The domains
$\Delta^+$ and $\Delta^-$ are unlimited(fig. 5.1).

Possible prove that:

1) if $\delta\in \Delta^+$, that $\arg G(+\infty)=2\pi$ (the curve
$\Lambda$ does one turn in positive direction comparatively begin
coordinates),

2) if $\delta\in \Delta^-$, that $\arg G(+\infty)=0$ (the curve
$\Lambda$ does not cover begin coordinates).

\begin{figure}[ht]
\begin{center}
\includegraphics{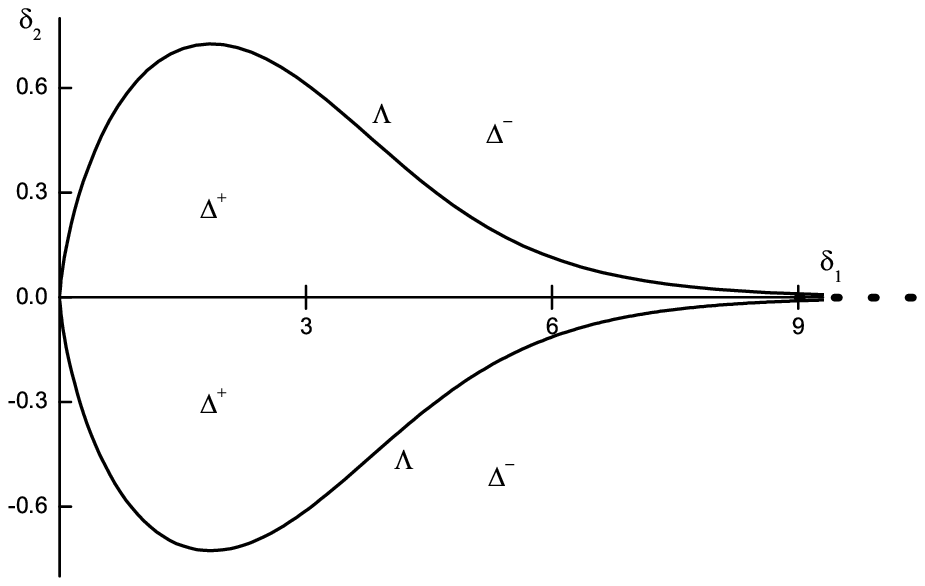}%[width=14cm,height=9cm]
\end{center}
\begin{center}
{Fig.1. The domain $\Delta^{\pm}$ on $\delta$--plane,
$\delta=(\delta_1,\delta_2)$.}
\end{center}
\end{figure}

According to (3.5) we get

1) if $\delta\in \Delta^+$, then $N=4$ -- dispersion function has
four zeroes, and

2) if $\delta\in \Delta^-$, then $N=2$  dispersion function has two
zeroes.

Go from $\delta$--planes to planes $(\omega_1,\nu_1)$, where
$\omega_1=\gamma/v_c$, $\nu_1=\varepsilon/v_c$, and
$\gamma=\omega/\omega_p$, $\varepsilon=\nu/\omega_p$. Find images
the unlimited domain $\Delta^{\pm}$ and curve $\Lambda$ under such
image.

\begin{figure}[ht]
\begin{center}
\includegraphics{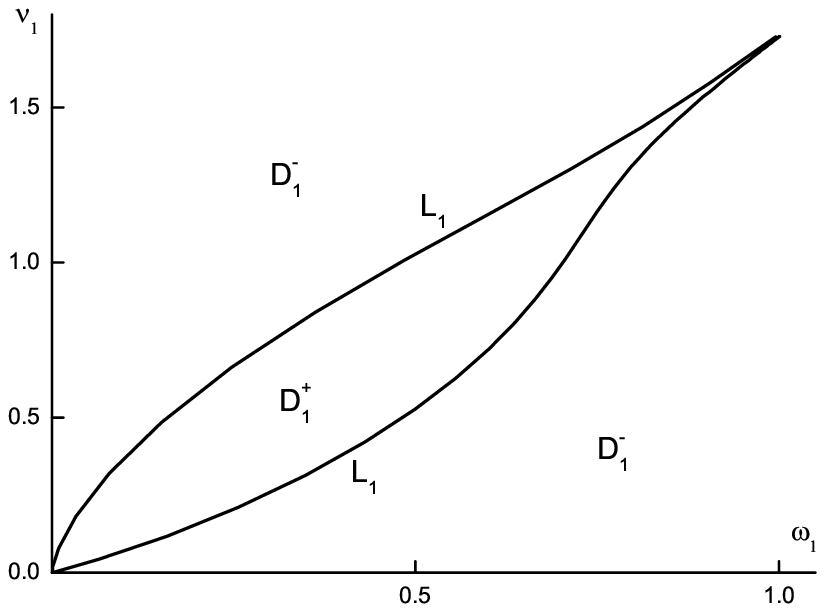}%[width=14cm,height=9cm]
\end{center}
\begin{center}
{Fig.2. Unlimited domain $D_1^{\pm}$ on $(\omega_1,\nu_1)$--plane.
Border of the domain $D_1^+$ is the curve $L_1=\partial D_1^+$.}
\end{center}
\end{figure}

Present the parametric equations crooked $\Lambda$ in form
$$
L_1: \quad -\omega_1^2+3\nu_1^2=p(\mu), \quad
-3\omega_1\nu_1+\dfrac{\nu_1^3}{\omega_1}=\pm q(\mu).
\eqno{(3.8)}
$$

From the first equation from (3.8) find:
$$
\omega_1=\pm \sqrt{3\nu_1^2-\,p(\mu)}. \eqno{(3.9)}
$$
Substitute (3.9) in the second equation from (3.8). We Get:
$$
\nu_1^3-3\nu_1(3\nu_1^2-p(\mu))= \pm
q(\mu)\sqrt{3\nu_1^2-p(\mu)}.
%\eqno{()}
$$
Involve this equation in square:
$$
64\nu_1^6-48\nu_1^4p(\mu)+3\nu_1^2 \big[
3p^2(\mu)-q^2(\mu)\big]+q^2(\mu)p(\mu)=0. \eqno{(3.10)}
$$

Under each fixed $\mu\in [0,+\infty]$ the equation (3.10) has a
single zero, which we mark through $Y(\mu)$. Ensemble of all such
zeroes forms is function $\nu_1=Y(\mu), \;0 \leqslant \mu \leqslant
+\infty$.

Hence we find the first parametric equation of curve $L_1$:
$$
\varepsilon=v_cY(\mu), \qquad 0 \leqslant \mu \leqslant
+\infty. \eqno{(3.11)}
$$

Substituting (3.11) in the second from equations (3.8), find:
$$
\omega_1=\sqrt{3Y^2(\mu)-p(\mu)}, \qquad 0 \leqslant \mu
\leqslant +\infty. \eqno{(3.12)}
$$

According to (3.11) and (3.12) we get parametric equations curve
$L_1$ on planes $(\omega_1,\nu_1)$:
$$
L_1:\;\; \omega_1=\sqrt{3\,Y^2(\mu)-p(\mu)}, \; \nu_1=\,Y(\mu),
\; 0 \leqslant \mu \leqslant +\infty.
%\eqno{(3.13)}
$$

Go from plane ($\omega_1,\nu_1$) to planes parameter
$(\gamma,\varepsilon)$. Present the parametric equations crooked
$\Lambda$ in form:
$$
-\gamma^2+3\varepsilon^2=v_c^2p(\mu), \quad
-3\gamma\varepsilon+\dfrac{\varepsilon^3}{\gamma}=\pm v_c
q(\mu). \eqno{(3.13)}
$$
where
$$
\gamma=\dfrac{\omega}{\omega_p}, \quad\qquad
\varepsilon=\dfrac{\varepsilon}{\omega_p}.
$$

From the first equation from (3.13) find:
$$
\gamma=\pm \sqrt{3\varepsilon^2-v_c^2\,p(\mu)}. \eqno{(3.14)}
$$
Substitute (3.14) in the second equation from (3.13). We Get:
$$
\varepsilon^3-3\varepsilon(3\varepsilon^2-v_c^2p(\mu))= \pm
v_c^2 q(\mu)\sqrt{3\varepsilon^2-v_c^2p(\mu)}.
%\eqno{()}
$$
Involve this equation in square:
$$
64\Big(\dfrac{\varepsilon}{v_c}\Big)^6-48\Big(\dfrac{\varepsilon}
{v_c}\Big)^4p(\mu)+3\Big(\dfrac{\varepsilon}{v_c}\Big)^2 \Big[
3p^2(\mu)-q^2(\mu)\Big]+q^2(\mu)p(\mu)=0. \eqno{(3.15)}
$$

Under each fixed  $\mu\in [0,+\infty]$ the equation (3.15) has a
single zero, which we mark through  $Y(\mu)$. Ensemble of all such
zeroes forms is function
$$
 \dfrac{\varepsilon}{v_c}=Y(\mu), \;\qquad
0 \leqslant \mu \leqslant +\infty.
$$

Consequently, we have found one parametric equation of curve $L$,
being image of curve $\Lambda$:
$$
\varepsilon=v_cY(\mu), \qquad 0 \leqslant \mu \leqslant
+\infty. \eqno{(3.16)}
$$

Substituting (3.16) in the second from equations (3.13), we find:
$$
\gamma=v_c\sqrt{3Y^2(\mu)-p(\mu)}, \qquad 0 \leqslant \mu
\leqslant +\infty. \eqno{(3.17)}
$$

According (3.16) and (3.17) on $(\gamma,\varepsilon)$--plane we have
the parametric equati\-ons of curve $L(v_c)$:
$$
L(v_c):\;\; \gamma=v_c\sqrt{3\,Y^2(\mu)-p(\mu)}, \;\quad
\varepsilon=v_c\,Y(\mu), \; 0 \leqslant \mu \leqslant +\infty.
%\eqno{(3.13)}
$$

\begin{figure}[ht]
\begin{center}
\includegraphics{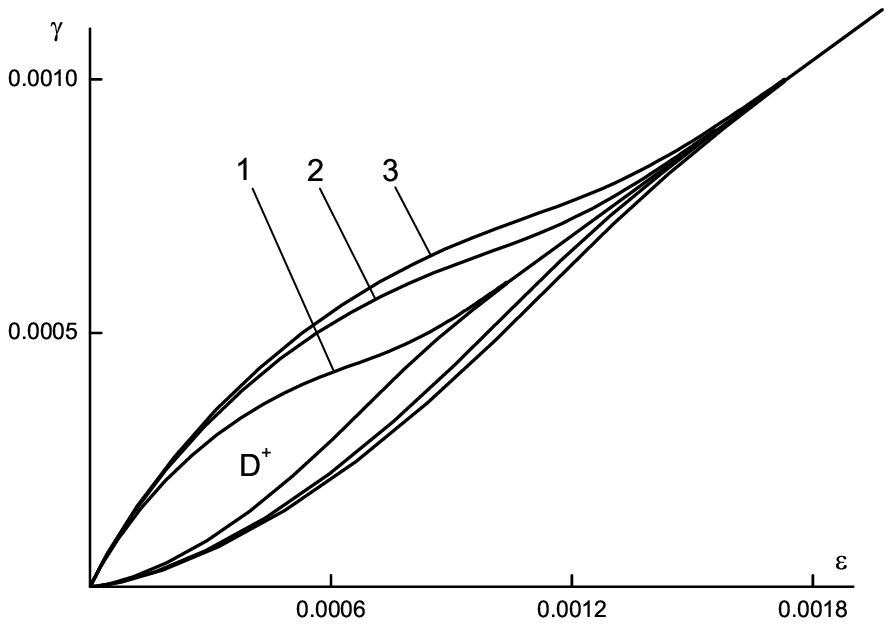}%[width=14cm,height=9cm]
\end{center}
\begin{center}
{ Fig. 3. The unlimited domain $D^{\pm}$ on the
$(\gamma,\varepsilon)$--plane. For curves $1,2,3$ $v_c~=
0.0006$\;$(T=1000^\circ K)$, \; $0.001$\;$(T=3000^\circ
K),$\;$0.013\;(T=5000^\circ K)$ respectively.}
\end{center}
\end{figure}

We find the general solution to system (1.3), (1.4) in the form of
an expansion in the eigenfunctions of the discrete and continuous
spectrum; this solution automatically satisfies the boundary
conditions at infinity:
$$
h(x,\mu)=\dfrac{a}{\sqrt{\pi}}\sum\limits_{k=0}^{1}
\dfrac{A_k\eta_k^3}{\eta_k-\mu}\exp\Big(-\eta_k^2-
\dfrac{z_0x}{\eta_k}\Big)+
$$
$$
+ \int\limits_{0}^{\infty}\exp\Big(-\dfrac{z_0x}{\eta}\Big)
A(\eta)\Phi(\eta,\mu)\,d\eta, \eqno{(2.10)}
$$
$$
e(x)=\dfrac{az_0}{\sqrt{\pi}}\sum\limits_{k=0}^{1}
A_k\eta_k^2\exp\Big(-\eta_k^2-\dfrac{z_0x}{\eta_k}\Big)+\hspace{3cm}
$$
$$\hspace{2cm}
+\dfrac{az_0}{\sqrt{\pi}}\int\limits_{0}^{\infty}
\exp\Big(-\eta^2-\dfrac{z_0x}{\eta}\Big)\eta^2\,A(\eta)\,d\eta.
\eqno{(2.11)}
$$

Here  $A_k (k=0,1)$ are the unknown coefficients corresponding to
the discrete spectrum,  $A_1=0$, if  $\delta\in D^-$; and $A(\eta)$
is an unknown function called the coefficient of the continuous
spectrum $\Re(z/\eta_k)>0 \quad (k=0,1)$, $\Re z_0=1$.

{\bf 4. The index of the problem equals zero.} We consider the case
when the problem index is equal to zero. Substituting expressions
(2.10) and (2.11) into (1.5), we obtain the singular integral
equation with the Cauchy kernel:
$$
\dfrac{a}{\sqrt{\pi}}\int\limits_{0}^{\infty}\dfrac{\eta^3
\exp(-\eta^2)A(\eta)}{\eta-\mu}\,d\eta+\lambda(\mu)A(\mu)+
a\varphi(\mu)=0, \quad 0<\mu<\infty, \eqno{(3.1)}
$$
where
$$
\varphi(\mu)=\dfrac{A_0}{\sqrt{\pi}}\dfrac{\eta_0^3\exp(-\eta_0^2)}
{\eta_0-\mu}.
$$

We define the auxiliary function in the complex plane
$$
N(z)=\dfrac{1}{\sqrt{\pi}}\int\limits_{0}^{\infty}
\dfrac{\eta^2\exp(-\eta^2)A(\eta)}{\eta-z}\,d\eta.
$$

Then the equation (3.1) is possible to transform
to the Riemann --- Hilbert
boundary value problem:
$$
\lambda^+(\mu)\Big[N^+(\mu)+\varphi(\mu)\Big]=
\lambda^-(\mu)\Big[N^-(\mu)+\varphi(\mu)\Big],\quad 0<\mu<\infty.
\eqno{(3.2)}
$$

We consider the Riemann --- Hilbert boundary value problem
on the half-axis:
$$
\dfrac{X^+(\mu)}{X^-(\mu)}=G(\mu),\qquad 0<\mu<+\infty.
$$
where
$$
G(\mu)=\dfrac{\lambda^+(\mu)}{\lambda^-(\mu)},
$$
$G(\mu)$ is the coefficient of the
boundary value problem.

The index in this problem is $\varkappa(G)=0$, then
the solution of
problem (3.2) has the form \cite{4}:
$$
X(z)=\exp V(z), \qquad V(z)=\dfrac{1}{2\pi i}
\int\limits_{0}^{\infty}\dfrac{\ln G(\tau)\,d\tau}{\tau-z}.
\eqno{(3.3)}
$$

We use (3.3) to reduce problem (3.2) for determining an analytic
function $N(z)$ from its zero jump problem on the cut:
$$
X^+(\mu)\Big[N^+(\mu)+\varphi(\mu)\Big]=
X^-(\mu)\Big[N^-(\mu)+\varphi(\mu)\Big],\quad 0<\mu<\infty.
\eqno{(3.4)}
$$

A general solution of (3.4) is given by the formula
$$
N(z)=\dfrac{A_0}{\sqrt{\pi}}\dfrac{\eta_0^3\exp(-\eta_0^2)}
{z-\eta_0}+\dfrac{C_0}{(z-\eta_0)X(z)}.\eqno{(3.5)}
$$
where  $C_0$ is an arbitrary constant.

Let's eliminate poles at the decision (3.5), we have:
$$
C_0=-\dfrac{A_0}{\sqrt{\pi}}\exp(-\eta_0^2)\eta_0^3X(\eta_0).
\eqno{(3.6)}
$$

Employing the Sokhotsky formula for difference from the boundary
va\-lu\-es of function  $N(z)$, we find the coefficient of the
continuous spectrum:
$$
2\sqrt{\pi}\,i\,\eta^3\,\exp(-\eta^2)\,A(\eta)=\dfrac{C_0}
{\eta-\eta_0}\Big[\dfrac{1}{X^+(\eta)}-\dfrac{1}{X^-(\eta)}\Big].
$$

From the definition of the auxiliary function  $N(z)$  we have:
$$
N(0)=\dfrac{1}{\sqrt{\pi}}\int\limits_0^{\infty}
\eta^2\exp(-\eta^2)A(\eta)d\eta.
\eqno{(3.7)}
$$

Considering equality (3.7), we write (2.11) in form:
$$
-\dfrac{C_0}{\eta_0X(\eta_0)}+N(0)=\dfrac{1}{az_0}.
$$

From general solution (3.5) the constant $C_0$  is found:
$$
C_0=-\dfrac{\eta_0}{az_0}X(0).
\eqno{(3.8)}
$$

From comparison (3.6) and (3.8) we get the unknown coefficients
corres\-pon\-ding to the discrete spectrum:
$$
A_0=\dfrac{\sqrt{\pi}\,X(0)}
{a\,z_0\,X(\eta_0)\,\eta_0^2\,\exp(-\eta_0^2)}.
$$

The derivative of expression (2.11) leads us to equality
$$
\dfrac{e'(x)}{az_0^2}=\dfrac{C_0}{\eta_0^2X(\eta_0)}
\exp(-\dfrac{z_0x}{\eta_0})-
\dfrac{1}{\sqrt{\pi}}\int\limits_0^\infty
\exp(-\eta^2-\dfrac{z_0x}{\eta})\eta
A(\eta)d\eta.
$$

From here we find the electric field derivative on plasma boundary
$x=0$:
$$
\dfrac{e'(0)}{az_0^2}=\dfrac{C_0}{\eta_0X(\eta_0)}-
\dfrac{1}{\sqrt{\pi}}\int\limits_0^\infty
\exp(-\eta^2)\eta A(\eta)d\eta.
\eqno{(3.9)}
$$

For calculation of integral from the expression (3.9) we shall
use a derivative of auxiliary function
$$
N'(\mu)=\dfrac{1}{\sqrt{\pi}}\int\limits_{0}^{\infty}
\dfrac{\eta^3\exp(-\eta^2)A(\eta)
\,d\eta} {(\eta-\mu)^2}.\eqno{(3.10)}
$$

Using the expression (3.10) we find the derivative in zero:
$$
N'(0)=\dfrac{1}{\sqrt{\pi}}\int\limits_{0}^{\infty}
\eta\exp(-\eta^2)A(\eta)
\,d\eta.\eqno{(3.11)}
$$

Using expression for coefficient of the continuous spectrum by
(3.11) we find:
$$
N'(z)=-\dfrac{C_0}{(z-\eta_0)^2}\Big(\dfrac{1}{X(z)}-
\dfrac{1}{X(\eta_0)}\Big)-\dfrac{C_0}{z-\eta_0}\dfrac{X'(z)}
{X^2(z)},
$$
whence
$$
N'(0)=-\dfrac{C_0}{\eta_0^2}\Big(\dfrac{1}{X(0)}-
\dfrac{1}{X(\eta_0)}\Big)-\dfrac{C_0}{\eta_0}
\dfrac{X'(0)} {X^2(0)}.
\eqno{(3.12)}
$$

Substituting (3.12) into (3.9) we obtain the equation:
$$
e'(0)=\dfrac{az_0^2C_0}{\eta_0X(0)}\bigg[\dfrac{1}{\eta_0}-
\dfrac{X'(0)}{X(0)}\bigg]=z_0\bigg[\dfrac{X'(0)}{X(0)}-
\dfrac{1}{\eta_0}\bigg]. \eqno{(3.13)}
$$

In \cite{2} the following formula for the impedance is given:
$$
Z=\dfrac{4\pi i \omega l}{c^2z_0}\cdot
\dfrac{e(0)}{e'(0)}.\eqno{(3.14)}
$$

Substituting (3.13) into (3.14), we obtain the
exact expression for
the impedance:
$$
Z=\dfrac{4\pi i \omega l}{c^2 \,z_0}\bigg[\dfrac{X'(0)}{X(0)}-
\dfrac{1}{\eta_0}\bigg]^{-1}. \eqno{(3.15)}
$$

The equation (3.15) is the exact expression for calculation
of the surface
impedance. This equation expresses value of the impedance in terms
of the function  $X(z)$ and zeros of  the
dispersion function of the problem.

{\bf 5. The index of the problem equals one.}
%{\bf 4. Problem index is equal one.}
Now we consider case when the
problem index is equal one, i.e.  $\varkappa(G)=1$.

Substituting expressions (2.10) and (2.11) into (1.5),
we obtain the
singular integral equation with the Cauchy kernel:
$$
a\varphi(\mu)+\dfrac{a}{\sqrt{\pi}}\int\limits_{0}^{\infty}
\dfrac{\eta^3\exp(-\eta^2)A(\eta)}{\eta-\mu}\,d\eta+
\lambda(\mu)A(\mu)=0, \quad 0<\mu<\infty, \eqno{(4.1)}
$$
where
$$
\varphi(\mu)=\dfrac{1}{\sqrt{\pi}}\sum\limits_{k=0}^{1}
\dfrac{A_k\eta_k^3\exp(-\eta_k^2)}{\eta_k-\mu}.
$$

We define the auxiliary function in the complex plane:
$$
N(z)=\dfrac{1}{\sqrt{\pi}}\int\limits_{0}^{\infty}
\dfrac{\eta^3\exp(-\eta^2)A(\eta)}{\eta-z}\,d\eta,
$$
for which boundary values from above and from below on the
valid axis formulas of Sokhotskii are carried out:
$$
N^+(\mu)-N^-(\mu)=2\sqrt{\pi}\,i\,\mu^3\exp(-\mu^2)A(\mu).
$$

Using the boundary values  $N^\pm(\mu)$ and
$\lambda^\pm(\mu)$, we
pass from the singular equation (4.1) to the
Riemann --- Hilbert boundary value
problem:
$$
\lambda^+(\mu)[N^+(\mu)+\varphi(\mu)]=\lambda^-(\mu)[N^-(\mu)+
\varphi(\mu)], \quad 0<\mu<\infty. \eqno{(4.2)}
$$

Let's solve a corresponding Riemann --- Hilbert
boundary value problem:
$$
\dfrac{X^+(\mu)}{X^-(\mu)}=G(\mu),\qquad 0<\mu<+\infty,
\eqno{(4.3)}
$$
where
$$
G(\mu)=\dfrac{\lambda^+(\mu)}{\lambda^-(\mu)},\qquad
0<\mu<\infty.
$$

Since problem index is equal one $\varkappa(G)=1$  as the solution
of (4.3) we take function
$$
X(z)=\dfrac{1}{z}\exp V(z),\quad V(z)=\dfrac{1}{2\pi
i}\int\limits_{0}^{\infty} \dfrac{\ln G(\tau)-2\pi
i}{\tau-z}\,d\tau. \eqno{(4.4)}
$$

Using (4.4) we transform the boundary value problem (4.2) to
determining an analytic function from its zero jump problem
on the cut:
$$
X^+(\mu)[N^+(\mu)+\varphi(\mu)]=X^-(\mu)[N^-(\mu)+ \varphi(\mu)],
\quad 0<\mu<\infty. \eqno{(4.5)}
$$

The general solution of (4.5) is following
$$
N(z)=\dfrac{1}{\sqrt{\pi}}\sum\limits_{k=0}^{1}
\dfrac{A_k\eta_k^3\exp(-\eta_k^2)}{z-\eta_k}+
\sum\limits_{k=0}^{1}\dfrac{C_k}{(z-\eta_k)X(z)}, \eqno{(4.6)}
$$
where  $C_0$, $C_1$  is the arbitrary values.

 Since $N(\infty)=0$,  from here we receive that $C_0+C_1=0$.

Let's eliminate poles at the decision (4.6), we get:
$$
C_0=-\dfrac{A_0}{\sqrt{\pi}}\eta_0^3\exp(-\eta_0^2)X(\eta_0),
\quad
 C_1=-\dfrac{A_1}{\sqrt{\pi}}\eta_1^3\exp(-\eta_1^2)X(\eta_1).
\eqno{(4.7)}
$$

Substituting the general solution (4.6) in the Sokhotskii formula
for (4.6) we find unknown function called the coefficient of the
continuous spectrum:
$$
\eta^3\exp(-\eta^2)A(\eta)=\dfrac{1}{2\sqrt{\pi}\,i}
\bigg[\dfrac{1}{X^+(\eta)}-\dfrac{1}{X^-(\eta)}\bigg]
\sum\limits_{k=0}^{1}\dfrac{C_k}{\eta-\eta_k}. \eqno{(4.8)}
$$

Using (4.7) we reduce (2.10) to the form:
$$
-\dfrac{C_0}{\eta_0X(\eta_0)}-\dfrac{C_1}{\eta_1X(\eta_1)}+
\dfrac{1}{\sqrt{\pi}}\int\limits_0^\infty
\eta^2\exp(-\eta^2)A(\eta)d\eta=
\dfrac{1}{az_0},
\eqno{(4.9)}
$$

From equation (4.9) we will find last integral
using definition of function
$N(z)$:
$$
N(0)=\dfrac{1}{\sqrt{\pi}}\int\limits_0^\infty\eta^2
\exp(-\eta^2)A(\eta)d\eta.
$$

Using the common decision (4.6), we found  $N(0)$:
$$
N(0)=C_0\left[\dfrac{1}{\eta_0X(\eta_0)}-
\dfrac{1}{\eta_1X(\eta_1)}-\dfrac{1}
{\eta_0X(0)}+\dfrac{1}{\eta_1X(0)}\right]
$$

We substitute this expression into  the equation (4.9):
$$
C_0=\dfrac{X(0)}{az_0\Big(\dfrac{1}{\eta_1}-
\dfrac{1}{\eta_0}\Big)}.
\eqno{(4.10)}
$$

Differentiating expansion for electric field (2.11) and using (4.7),
where $x=0$  we found:
$$
e'(0)=\dfrac{C_0az_0^2}{\eta_0^2X(\eta_0)}-
\dfrac{C_0az_0^2}{\eta_1^2X(\eta_1)}-\dfrac{az_0^2}{\sqrt{\pi}}
\int\limits_{0}^{\infty}\eta \exp(-\eta^2)A(\eta)\,d\eta.
\eqno{(4.11)}
$$

Integral from (4.11) is the auxiliary function derivative
$N'(0)$:
$$
N'(0)=\dfrac{1}{\sqrt{\pi}} \int\limits_{0}^{\infty}\eta
\exp(-\eta^2)A(\eta)\,d\eta.
$$

The common solution (4.6) we present in form:
$$
N(z)=\dfrac{1}{z-\eta_0}\bigg[\dfrac{C_0}{X(z)}+
\dfrac{1}{\sqrt{\pi}}A_0\eta_0^3\exp(-\eta_0^2)\bigg]+
$$
$$
+\dfrac{1}{z-\eta_1}\bigg[\dfrac{C_1}{X(z)}+
\dfrac{1}{\sqrt{\pi}}A_1\eta_1^3\exp(-\eta_1^2)\bigg].
$$

Derivative of this expression is equal:
$$
N'(z)=-\dfrac{1}{(z-\eta_0)^2}
\bigg[\dfrac{1}{X(z)}-\dfrac{1}{X(\eta_0)}\bigg]-
\dfrac{C_0}{z-\eta_0}
\dfrac{X'(z)}{X^2(z)}+
$$
$$
+\dfrac{C_0}{(z-\eta_1)^2}\bigg[\dfrac{1}{X(z)}-
\dfrac{1}{X(\eta_1)}\bigg]+\dfrac{C_0}{z-\eta_1}\dfrac{X'(z)}
{X^2(z)}.
$$

Hence, we obtain:
$$
N'(0)=-\dfrac{1}{\eta_0^2}
\bigg[\dfrac{1}{X(0)}-\dfrac{1}{X(\eta_0)}\bigg]+
\dfrac{C_0}{\eta_0}
\dfrac{X'(0)}{X^2(0)}+
$$
$$
+\dfrac{C_0}{\eta_1^2}\bigg[\dfrac{1}{X(0)}-
\dfrac{1}{X(\eta_1)}\bigg]-\dfrac{C_0}{\eta_1}\dfrac{X'(0)}
{X^2(0)}.
$$

Using this equality for the electric field derivative we have:
$$
e'(0)=\dfrac{C_0az_0^2}{X(0)}\Bigg[\bigg(\dfrac{1}{\eta_0}-
\dfrac{1}{\eta_1}\bigg)\bigg(\dfrac{1}{\eta_0}+
\dfrac{1}{\eta_1}\bigg)+\dfrac{X'(0)}{X(0)}\bigg(\dfrac{1}{\eta_0}-
\dfrac{1}{\eta_1}\bigg)\Bigg].
$$

Therefore according to (4.10) we receive that
$$
e'(0)=z_0\Bigg[\dfrac{X'(0)}{X(0)}-\dfrac{1}{\eta_0}-
\dfrac{1}{\eta_1}\Bigg].
$$

Hence, expression for the impedance is equal:
$$
Z=\dfrac{4\pi i \omega l}{c^2 z_0}\cdot
\Bigg[\dfrac{X'(0)}{X(0)}-\dfrac{1}{\eta_0}-
\dfrac{1}{\eta_1}\Bigg]^{-1}.
$$

This equation expresses value of the impedance in terms of factor
function $X(z)$  and zeros of dispersive function of the problem.

\begin{center}
  CONCLUSIONS
\end{center}

A closed form solution of a system of two equations Boltzmann ---
Vlasov and Maxwell arising in the skin effect problem for a
Maxwillian plasmas is presented. The kinetic Boltzmann --- Vlasov
equation with a $\tau$--model collision operator is considered.
Case's \cite{7} method and Riemann --- Hilbert boundary value
problem \cite{8} with the coefficient $G(\mu)=\dfrac{\lambda^+(\mu)}
{\lambda^-(\mu)}$, where $\lambda(z)$ is the dispersion function of
the problem. The discrete fashion is obtained. Consider the domain
$D^+$, that if the frequency lies in this domain, there are four
discrete solution, and if the frequency is outside this domain,
there are two discrete solutions. The exact solution of the initial
boundary value problem with diffusion scattering of electrons from
plasma boundary is constructed. The exact formula for calculation of
the surface impedance is obtained.


\begin{thebibliography}{18}
\bibitem{1} {\sl A.F. Alexandrov, I.S. Bogdankevich and
A.A. Rukhadze.} Principles of Plasma Electrodynamics
(Springer--Verlag, New York, 1984).
%\bibitem{2}{\sl Silin V.P., Rukhadze A.A.}
%Electromagnetic properties of plasma and mediums like plasma.
%\cite{}Atomizdat, Moscow, 1961. P. 244. (in Russian).

\bibitem{2}{\sl A.V. Latyshev and A.A. Yushkanov.}
Analytical solutions in the skin
effect theory. Monography.
Moscow State Regional University, Moscow, 2008. P. 285. (in Russian).


\bibitem{3}{\sl A.V. Latyshev and A.A. Yushkanov.}
Analytical solutions of
the boundary problem of the kinetic theory. Monography.
Moscow State Regional University, Moscow, 2004. P. 263. (in Russian).

\bibitem{4} {\sl M. Opher,  G.J. Morales and J.N. Leboeuf.}
Krook collisional models of the kinetic susceptibility of
plasmas// Phys. Rev. E. 2002 {\bf 66}(1), 016407, pp.  66 -- 75.

\bibitem{5}{\sl I.D. Kaganovich, O.V. Polomarov and C.E.
Theodosiou.}
Resisting the anomalous  rf  field penetration into a warm
plasma// ArXiv: physics/0506135.

\bibitem{6}{\sl M. Dressel and G. Gr\"{u}ner }  Electrodynamics of Solids. Optical
Properties of Electrons in Matter. Cambridge university press. 2002.
P. 474.


\bibitem{7} {\sl V.S. Vladimirov and  V.V. Zharinov.}
Equations of mathematical physics.
Fizmatlit, Moscow, 2001. P. 400. (in Russian).

\bibitem{8} {\sl K.M. Case and P.M. Zweifel.} Linear
Transport Theory. Addison -- Wesley. 1967.

\bibitem{9} {\sl F.D. Gakhov.} Boundary -- Value Problems [in
Russian]. Nauka. Moscow. 1977.

\bibitem{13} {\sl Zimbovskay N.A.} Fermi--liquid theory of the surface impedance of
 a metal in a normal magnetic field (2006) Phys. Rev. B \textbf{74}
035110.

\bibitem{14} {\sl Zimbovskay N.A.} Local geometry of the Fermi surface
and the skin effect in layered conductors (1998) JETP \textbf{86} 6.

\end{thebibliography}
\end{document}